# THE QUANTUM WAVE PACKET AND THE FEYNMAN–DE BROGLIE–BOHM PROPAGATOR OF THE SCHRODINGER–NASSAR EQUATION FOR AN EXTENDED ELECTRON

J. M. F. Bassalo[1], P. T. S. Alencar[2], D. G. da Silva[3], A. B. Nassar[4] and M. Cattani[5]

[1]Fundação Minerva,Avenida Governador José Malcher 629 - CEP 66035-100, Belém, Pará, Brasil
E−mail:jmfbassalo@gmail.com

[2]Universidade Federal do Pará - CEP 66075-900, Guamá, Belém, Pará, Brasil
E−mail:tarso@ufpa.br

[3]Escola Munguba do Jari, Vitória do Jari - CEP 68924-000, Amapá, Brasil
E−mail: danielgemaque@yahoo.com.br

[4]Extension Program-Department of Sciences,University of California,Los Angeles,California 90024, USA
E−mail: nassar@ucla.edu/

[5]Instituto de Fìsica da Universidade de São Paulo. C. P. 66318, CEP 05315-970, São Paulo, SP, Brasil
E−mail: mcattani@if.usp.br

Abstract: In this paper we study the quantum wave packet and the Feynman−de Broglie−Bohm Propagator of the Schrodinger−Nassarequation for an extended electron. PACS 03.65 − Quantum Mechanics

1. **Introduction**

In the present work we investigate the quantum wave packet and the Feynman Propagator of the Schrodinger equation for an extended electron proposed by Nassar, [1] using the Quantum Mechanical of the de Broglie−Bohm.[2].

**2. The Schrodinger−Nassar Equation for an Extended Electron**

About a century ago, Lorentz [3] and Abraham [4] argued that when an electron (with velocity $v$ and charge $e$), is accelerated, there are additional forces acting due to the electron's own electromagnetic field. The so-called Lorentz-Abraham equation for a point-charge electron:

$$m \, dv/dt = (2 \, e^2)/(3 \, c^2) \, d^2v/dt^2 + F_{ext}. \quad (2.1)$$

was found to be unsatisfactory because, for $F_{ext} = 0$, it admits runaway solutions. These solutions clearly violate the law of inertia.

Since the seminal works of Lorentz and Abraham, many papers and textbooks have given great consideration to the proper equation of motion of an electron. [5]–[13] The problematic runaway solutions were solved by Sommerfeld [7] and Page [8] adopting an *extended model*. In the non–relativistic case of a sphere with uniform surface charge, such an electron obeys in good approximation the difference-differential equation: [9]–[11]

$$m\, dv/dt = (e^2)/(3\, L^2\, c)\, [v(t - 2\, L/c) - v(t)] + F_{ext}. \quad (2.2)$$

This extended model is finite and causal if the electron size $L$ is larger than the classical electron radius $r_e = e^2/(m\, c^2)$. We will analyze here only the case of the sphere with uniform surface charge; the case of a volume charge is considerably more complicated and adds nothing to the understanding of the problem.

The dynamics of charges is a key example of the importance of obeying the validity limits of a physical theory. If classical equations can no longer be trusted at distances of the order (or below) the Compton wavelength, what is the Schrodinger equation that can replace equation (2.2)? Within *QED*, workers have not been able to derive an equation of motion and it is unclear whether *QED* can actually produce an equation of motion at all.

By using the quantum mechanical approach of de Broglie–Bohm,[2] Nassar proposed [1] an answer to this problem in the non relativistic regime. Nassar [1] have proposed the following equation for an extended electron, that will named Schrodinger–Nassar Equation (SNEEE) :

$$i\, \hbar\, \frac{\partial \psi(x, t)}{\partial t} = -\frac{\hbar^2}{2\, m}\, \frac{\partial^2 \psi(x, t)}{\partial x^2} + (i\, \hbar\, \alpha\, \ell n\, [\, \frac{\psi(x, t - 2\, L/c)\, \psi^*(x, t)}{\psi^*(x, t - 2\, L/c)\, \psi(x, t)}\, ])\, \psi(x, t)\,, \quad (2.3a)$$

where:

$$\alpha = \frac{e^2}{6\, m\, L^2\, c}, \quad (2.3b)$$

is a constant, $\psi(x, t)$ and $\psi(x, t - 2\, L/c)$ the extended electron wave functions.

Now, let us obtain the quantum wave packet for (2.3a) for an extended electron. Initially, writing the wavefunction $\psi(x, t)$ and $\psi(x, t - 2\, L/c)$ in polar form, defined by the Madelung–Bohm transformation, [14, 15] we get:

$$\psi(x, t) = \phi(x, t)\, e^{i\, S(x, t)}\,, \quad (2.4a)$$

$$\psi(x, t - 2\, L/c) = \phi(x, t)\, e^{i\, S(x, t - 2\, L/c)}\,, \quad (2.4b)$$

where $S(x, t)$ is the classical action and $\phi(x, t)$ will be defined in what follows.

Calculating the derivatives, temporal and spatial, of (2.4a), we get:

$$\frac{\partial \psi(x,t)}{\partial t} = e^{i S(x,t)} \frac{\partial \phi(x,t)}{\partial t} + i \phi(x,t) e^{i S(x,t)} \frac{\partial S(x,t)}{\partial t} \rightarrow$$

$$\frac{\partial \psi}{\partial t} = e^{i S} \left( \frac{\partial \phi}{\partial t} + i \phi \frac{\partial S}{\partial t} \right) = \left( i \frac{\partial S}{\partial t} + \frac{1}{\phi} \frac{\partial \phi}{\partial t} \right) \psi, \quad (2.5\text{a,b})$$

$$\frac{\partial \psi}{\partial x} = e^{i S} \left( \frac{\partial \phi}{\partial x} + i \phi \frac{\partial S}{\partial x} \right) = \left( i \frac{\partial S}{\partial x} + \frac{1}{\phi} \frac{\partial \phi}{\partial x} \right) \psi, \quad (2.5\text{c,d})$$

$$\frac{\partial^2 \psi}{\partial x^2} = \frac{\partial}{\partial x} \left[ \left( i \frac{\partial S}{\partial x} + \frac{1}{\phi} \frac{\partial \phi}{\partial x} \right) \psi \right] =$$

$$= \psi \left[ i \frac{\partial^2 S}{\partial x^2} + \frac{1}{\phi} \frac{\partial^2 \phi}{\partial x^2} - \frac{1}{\phi^2} \left( \frac{\partial \phi}{\partial x} \right)^2 \right] + \left( i \frac{\partial S}{\partial x} + \frac{1}{\phi} \frac{\partial \phi}{\partial x} \right) \frac{\partial \psi}{\partial x} =$$

$$= \psi \left[ i \frac{\partial^2 S}{\partial x^2} + \frac{1}{\phi} \frac{\partial^2 \phi}{\partial x^2} - \frac{1}{\phi^2} \left( \frac{\partial \phi}{\partial x} \right)^2 \right] + \left( i \frac{\partial S}{\partial x} + \frac{1}{\phi} \frac{\partial \phi}{\partial x} \right) \left( i \frac{\partial S}{\partial x} + \frac{1}{\phi} \frac{\partial \phi}{\partial x} \right) \psi =$$

$$= \psi \left[ i \frac{\partial^2 S}{\partial x^2} + \frac{1}{\phi} \frac{\partial^2 \phi}{\partial x^2} - \frac{1}{\phi^2} \left( \frac{\partial \phi}{\partial x} \right)^2 - \left( \frac{\partial S}{\partial x} \right)^2 + \frac{1}{\phi^2} \left( \frac{\partial \phi}{\partial x} \right)^2 + 2 \frac{i}{\phi} \frac{\partial S}{\partial x} \frac{\partial \phi}{\partial x} \right] \rightarrow$$

$$\frac{\partial^2 \psi}{\partial x^2} = e^{i S} \left[ \frac{\partial^2 \phi}{\partial x^2} + 2 i \frac{\partial S}{\partial x} \frac{\partial \phi}{\partial x} + i \phi \frac{\partial^2 S}{\partial x^2} - \phi \left( \frac{\partial S}{\partial x} \right)^2 \right] \quad (\div \phi) \rightarrow$$

$$\frac{\partial^2 \psi}{\partial x^2} = \left[ i \frac{\partial^2 S}{\partial x^2} + \frac{1}{\phi} \frac{\partial^2 \phi}{\partial x^2} - \left( \frac{\partial S}{\partial x} \right)^2 + 2 i \frac{1}{\phi} \frac{\partial S}{\partial x} \frac{\partial \phi}{\partial x} \right] \psi. \quad (2.5\text{e})$$

Now, inserting the relations defined by (2.4a,b; 2.5a,e) into (2.3a) we have [remembering that $e^{i S}$ is common factor and $\ln(x.y) = \ln x + \ln y$]:

$$i \hbar \left( \frac{\partial \phi}{\partial t} + i \phi \frac{\partial S}{\partial t} \right) = - \frac{\hbar^2}{2m} \left[ \frac{\partial^2 \phi}{\partial x^2} + \right.$$

$$\left. + 2 i \frac{\partial S}{\partial x} \frac{\partial \phi}{\partial x} + i \phi \frac{\partial^2 S}{\partial x^2} - \phi \left( \frac{\partial S}{\partial x} \right)^2 \right] +$$

$$+ \left( i \hbar \alpha \ln \left[ \frac{\phi(x,t) e^{i S(x, t - 2L/c)} \phi(x,t) e^{-i S(x,t)}}{\phi(x,t) e^{-i S(x, t - 2L/c)} \phi(x,t) e^{i S(x,t)}} \right] \right) \phi(x,t) \rightarrow$$

$$i \hbar \left( \frac{\partial \phi}{\partial t} + i \phi \frac{\partial S}{\partial t} \right) = - \frac{\hbar^2}{2m} \left[ \frac{\partial^2 \phi}{\partial x^2} + \right.$$

$$\left. + 2 i \frac{\partial S}{\partial x} \frac{\partial \phi}{\partial x} + i \phi \frac{\partial^2 S}{\partial x^2} - \phi \left( \frac{\partial S}{\partial x} \right)^2 \right] +$$

$$+ \left( i \hbar \alpha \ln \left[ \frac{e^{i S(x, t - 2L/c)} e^{-i S(x,t)}}{e^{-i S(x, t - 2L/c)} e^{i S(x,t)}} \right] \right) \phi(x,t) \rightarrow$$

$$i\hbar\left(\frac{\partial\phi}{\partial t}+i\phi\frac{\partial S}{\partial t}\right)=-\frac{\hbar^2}{2m}\left[\frac{\partial^2\phi}{\partial x^2}+\right.$$

$$\left.+2i\frac{\partial S}{\partial x}\frac{\partial\phi}{\partial x}+i\phi\frac{\partial^2 S}{\partial x^2}-\phi\left(\frac{\partial S}{\partial x}\right)^2\right]+$$

$$+(i\hbar\alpha\,\ell n\,[\,\frac{e^{i\,S(x,\,t-2L/c)}\,e^{i\,S(x,\,t-2L/c)}}{e^{i\,S(x,\,t)}\,e^{i\,S(x,\,t)}}\,])\,\phi(x,t)\quad\rightarrow$$

$$i\hbar\left(\frac{\partial\phi}{\partial t}+i\phi\frac{\partial S}{\partial t}\right)=-\frac{\hbar^2}{2m}\left[\frac{\partial^2\phi}{\partial x^2}+\right.$$

$$\left.+2i\frac{\partial S}{\partial x}\frac{\partial\phi}{\partial x}+i\phi\frac{\partial^2 S}{\partial x^2}-\phi\left(\frac{\partial S}{\partial x}\right)^2\right]+$$

$$+(i\hbar\alpha\,\ell n\,[\,\frac{e^{2i\,S(x,\,t-2L/c)}}{e^{2i\,S(x,\,t)}}\,])\,\phi(x,t)\quad\rightarrow$$

$$i\hbar\left(\frac{\partial\phi}{\partial t}+i\phi\frac{\partial S}{\partial t}\right)=-\frac{\hbar^2}{2m}\left[\frac{\partial^2\phi}{\partial x^2}+\right.$$

$$\left.+2i\frac{\partial S}{\partial x}\frac{\partial\phi}{\partial x}+i\phi\frac{\partial^2 S}{\partial x^2}-\phi\left(\frac{\partial S}{\partial x}\right)^2\right]+$$

$$+(i\hbar\alpha\,[\,2i\,S(x,\,t-2L/c)-2i\,S(x,\,t)\,])\,\phi(x,t). \quad (2.6)$$

Taking the real and imaginary parts of (2.6), we obtain:
(a) <u>imaginary part</u>

$$\frac{\partial\phi(x,t)}{\partial t}=-\frac{\hbar}{2m}\left[\,2\,\frac{\partial S(x,t)}{\partial x}\frac{\partial\phi(x,t)}{\partial x}+\phi(x,t)\frac{\partial^2 S(x,t)}{\partial x^2}\,\right], \quad (2.7)$$

(b) <u>real part</u>

$$-\hbar\,\phi(x,t)\frac{\partial S(x,t)}{\partial t}=-\frac{\hbar^2}{2m}\left[\frac{\partial^2\phi(x,t)}{\partial x^2}-\phi(x,t)\left[\frac{\partial S(x,t)}{\partial x}\right]^2\right]+$$

$$+(2\hbar\alpha\,[\,S(x,t)-S(x,t-2L/c)\,])\,\phi(x,t)\quad(\div m\,\phi)\rightarrow$$

$$-\frac{\hbar}{m}\frac{\partial S(x,t)}{\partial t}=-\frac{\hbar^2}{2m^2\phi}\left(\frac{\partial^2\phi(x,t)}{\partial x^2}-\phi(x,t)\left[\frac{\partial S(x,t)}{\partial x}\right]^2\right)+$$

$$+\left(\frac{2\hbar\alpha}{m}\,[\,S(x,t)-S(x,t-2L/c)\,]\right)^{\cdot} \quad (2.8)$$

## 2.1 Dynamics of the SNE for an Extended Electron

Now, let us see the correlation between (2.7–8) and the traditional equations of the Ideal Fluid Dynamics: [16] (a) continuity equation and (b) Euler's equation. To do this let us perform the following correspondences:

Quantum density probability:   $|\psi(x, t)|^2 = \psi^*(x, t)\,\psi(x, t)$

Quantum mass density:   $\rho(x, t) = \phi^2(x, t)$   $\sqrt{\rho} = \phi$,   (2.9a,b)

Gradient of the wave function:   $\dfrac{\hbar}{m}\dfrac{\partial S(x, t)}{\partial x}$

Quantum velocity:   $v_{qu}(x, t) \equiv v_{qu}$,   (2.10a)

Gradient of the wave function extended:   $\dfrac{\hbar}{m}\dfrac{\partial S(x, t - 2L/c)}{\partial x}$

Quantum velocity extended:   $v_{qu}(x, t - 2L/c) \equiv v_{que}$.   (2.10b)

Putting (2.9b, 2.10a) into (2.7) we get:

$$\frac{\partial \sqrt{\rho}}{\partial t} = -\frac{\hbar}{2m}\left(2\frac{\partial S}{\partial x}\frac{\partial \sqrt{\rho}}{\partial x} + \sqrt{\rho}\frac{\partial^2 S}{\partial x^2}\right) \rightarrow$$

$$\frac{1}{2\sqrt{\rho}}\frac{\partial \rho}{\partial t} = -\frac{\hbar}{2m}\left(2\frac{\partial S}{\partial x}\frac{1}{2\sqrt{\rho}}\frac{\partial \rho}{\partial x} + \sqrt{\rho}\frac{\partial^2 S}{\partial x^2}\right) \rightarrow$$

$$\frac{1}{\rho}\frac{\partial \rho}{\partial t} = -\frac{\hbar}{m}\left(\frac{\partial S}{\partial x}\frac{1}{\rho}\frac{\partial \rho}{\partial x} + \frac{\partial^2 S}{\partial x^2}\right) \rightarrow$$

$$\frac{1}{\rho}\frac{\partial \rho}{\partial t} = -\frac{\partial}{\partial x}\left(\frac{\hbar}{m}\frac{\partial S}{\partial x}\right) - \frac{1}{\rho}\left(\frac{\hbar}{m}\frac{\partial S}{\partial x}\right)\frac{\partial \rho}{\partial x} \rightarrow$$

$$\frac{1}{\rho}\frac{\partial \rho}{\partial t} = -\frac{\partial v_{qu}}{\partial x} - \frac{v_{qu}}{\rho}\frac{\partial \rho}{\partial x} \rightarrow$$

$$\frac{\partial \rho}{\partial t} + \rho\frac{\partial v_{qu}}{\partial x} + v_{qu}\frac{\partial \rho}{\partial x} = 0 \rightarrow$$

$$\frac{\partial \rho}{\partial t} + \frac{\partial(\rho\, v_{qu})}{\partial x} = 0,\quad (2.11)$$

which represents the continuity equation of the mass conservation law of the Fluid Dynamics. We must note that this expression also indicates descoerence of the considered physical system represented by (2.3a). Using (2.9b) we now define the quantum potential $V_{qu}$:

$$V_{qu}(x, t) \equiv V_{qu} = -\left(\frac{\hbar^2}{2m\phi}\right)\frac{\partial^2 \phi}{\partial x^2} = -\frac{\hbar^2}{2m}\frac{1}{\sqrt{\rho}}\frac{\partial^2 \sqrt{\rho}}{\partial x^2},\quad (2.12\text{a,b})$$

and (2.8) will written as:

$$\frac{\hbar}{m}\frac{\partial S}{\partial t} + \frac{\hbar^2}{2m^2}\left(\frac{\partial S}{\partial x}\right)^2 = -\left(\frac{2\hbar\alpha}{m}[S(x, t) - S(x, t - 2L/c)]\right) - \frac{V_{qu}}{m},\quad (2.13)$$

or, using (2.10a) and multiplying by $m$:

$$\hbar \frac{\partial S}{\partial t} + \frac{1}{2} m v_{qu}^2 + ( 2 \hbar \alpha [ S(x, t) - S(x, t - 2L/c) ] ) + V_{qu} \rightarrow \quad (2.14a)$$

$$\hbar \frac{\partial S}{\partial t} + \frac{1}{2} m v_{qu}^2 + V_{ee} + V_{qu} = 0, \quad (2.14b)$$

where:

$$V_{ee} = 2 \hbar \alpha [ S(x, t) - S(x, t - 2L/c) ], \quad (2.14c)$$

is the *potential extended*.

Differentiating (2.14a) with respect to $x$ and using (2.10a,b) we obtain:

$$\frac{\partial}{\partial x} [ \hbar \frac{\partial S}{\partial t} + \frac{1}{2} m v_{qu}^2 +$$
$$+ ( 2 \hbar \alpha [ S(x, t) - S(x, t - 2L/c) ] ) + V_{qu} ] = 0 \quad (\div m) \rightarrow$$

$$\frac{\partial}{\partial t}( \frac{\hbar}{m} \frac{\partial S}{\partial x}) + \frac{\partial}{\partial x} ( \frac{1}{2} v_{qu}^2 ) + 2 \alpha (v_{qu} - v_{que}) = - \frac{1}{m} \frac{\partial}{\partial x} V_{qu} \rightarrow$$

$$\frac{\partial v_{qu}}{\partial t} + v_{qu} \frac{\partial v_{qu}}{\partial x} = 2 \alpha [v_{qu}(t- 2 L/c) - v_{qu}(t)] - \frac{1}{m} \frac{\partial}{\partial x} V_{qu}. \quad (2.15)$$

Considering the "substantive differentiation" (local plus convective) or "hidrodynamic differentiation" [16]:

$$\frac{d}{dt} = \frac{\partial}{\partial t} + v_{qu} \frac{\partial}{\partial x}, \quad (2.16a)$$

$$v_{qu}(t) = \frac{dx_{qu}}{dt}, \quad (2.16b)$$

(2.15) becomes:

$$\frac{d v_{qu}}{dt} = 2 \alpha [v_{qu}(t- 2 L/c) - v_{qu}(t)] - \frac{1}{m} \frac{\partial}{\partial x} V_{qu}. \quad (2.17)$$

Multiplying the eq. (2.17) by $m$ and considering (2.3b), we have:

$$m \frac{d v_{qu}}{dt} = (e^2)/(3 L^2 c) [v_{qu}(t- 2 L/c) - v_{qu}(t)] + F_{qu}, \quad (2.18a)$$

where:

$$F_{qu} = - \frac{\partial}{\partial x} V_{qu}. \quad (2.18b)$$

is the *quantum force*. We note that (2.18a) is the quantum form of the Sommerfeld–Page's equation [see (2.2)] for a spherical electron with uniform surface charge density in the absence of external forces.

In what follows we calculate the wave packet of the *SNEEE* given by (2.3a).

**3. The Quantum Wave Packet of the Schrodinger Equation for an Extended Electron**

In order to find the quantum wave packet of the *SNEEE* the following *ansatz* is made:[1]

$$\rho(x, t) = [2\pi a^2(t)]^{-1/2} \exp\left(-\frac{[x - q(t)]^2}{2 a^2(t)}\right), \quad (3.1)$$

where $a(t)$ and $q(t)$ are auxiliary functions of time, to will be determined in what follows; representing the *width* and *center of mass of wave packet*, respectively.

Substituting (3.1) into (2.11) and integrated the result, we obtain:[1,17]

$$v_{qu}(x, t) = \frac{\dot{a}(t)}{a(t)}[x - q(t)] + \dot{q}(t), \quad (3.2)$$

where the integration constant must be equal to zero since $\rho$ and $\rho\, \partial S/\partial x$ vanish for $|x| \to \infty$. In fact, any well-behaved function of $(x - X)$ multiplied by $\rho$ clearly vanishes as $|x| \to \infty$.

To obtain the quantum wave packet of the *NSEEE* given by (2.3a,b), let us expand the functions $S(x, t)$, $S(x, t - 2L/c)$ and $V_{qu}(x\,t)$ around of $q(t)$ up to second Taylor order. In this way we have:

$$S(x, t) = S[q(t), t] + S'[q(t), t][x - q(t)] + \frac{S''[q(t), t]}{2}[x - q(t)]^2, \quad (3.3a)$$

$$S(x, t - 2L/c) = S[q(t), t - 2L/c] + S'[q(t), t - 2L/c][x - q(t)] +$$

$$+ \frac{S''[q(t), t - 2L/c]}{2}[x - q(t)]^2, \quad (3.3b)$$

$$V_{qu}(x, t) = V_{qu}[q(t), t] + V_{qu}'[q(t), t][x - q(t)] + \frac{V_{qu}''[q(t), t]}{2}[x - q(t)]^2. \quad (3.4)$$

Differentiating (3.3a,b) with respect to *x*, multiplying the result by $\frac{\hbar}{m}$, using (2.10a,b) and (3.2) and taking into account the polynomial identity property, we obtain:

$$\frac{\hbar}{m} \frac{\partial S(x, t)}{\partial x} = \frac{\hbar}{m} ( S'[q(t), t] + S''[q(t), t] [x - q(t)] ) =$$

$$= v_{qu}(x, t) = [\frac{\dot{a}(t)}{a(t)}] [x_{qu} - q(t)] + \dot{q}(t) \quad \rightarrow$$

$$S'[q(t), t] = \frac{m \dot{q}(t)}{\hbar}, \quad S''[q(t), t] = \frac{m}{\hbar} \frac{\dot{a}(t)}{a(t)}, \quad (3.5a,b)$$

$$\frac{\hbar}{m} \frac{\partial S(x, t - 2L/c)}{\partial x} = \frac{\hbar}{m} ( S'[q(t), t - 2L/c] + S''[q(t), t - 2L/c] [x - q(t)] ) =$$

$$= v_{que}(x, t - 2L/c) = \frac{\dot{a}(t - 2L/c)}{a(t)} [x_{qu} - q(t)] + \dot{q}(t - 2L/c) \quad \rightarrow$$

$$S'[q(t), t - 2L/c] = \frac{m \dot{q}(t - 2L/c)}{\hbar}, \quad S''[q(t), t - 2L/c] = \frac{m}{\hbar} \frac{\dot{a}(t - 2L/c)}{a(t)}. \quad (3.5c,d)$$

Substituting (3.5a-d) into (3.3a,b), results:

$$S(x, t) = S_o(t) + \frac{m \dot{q}(t)}{\hbar} [x - q(t)] + \frac{m}{2\hbar} \frac{\dot{a}(t)}{a(t)} [x - q(t)]^2, \quad (3.6a)$$

$$S(x, t - 2L/c) = S_o(t - 2L/c) + \frac{m \dot{q}(t - 2L/c)}{\hbar} [x - q(t)] +$$

$$+ \frac{m}{2\hbar} \frac{\dot{a}(t - 2L/c)}{a(t)} [x - q(t)]^2, \quad (3.6b)$$

where:

$$S_o(t) \equiv S[q(t), t], \quad (3.7a)$$

$$S_o(t - 2L/c) \equiv S[q(t), t - 2L/c], \quad (3.7b)$$

are the classical actions.

Differentiating the (3.6a) with respect to $t$, we obtain (remembering that $\frac{\partial x}{\partial t} = 0$):

$$\frac{\partial S}{\partial t} = \dot{S}_o(t) + \frac{\partial}{\partial t}(\frac{m \dot{q}(t)}{\hbar} [x - q(t)]) + \frac{\partial}{\partial t}(\frac{m}{2\hbar} [\frac{\dot{a}(t)}{a(t)}] [x - q(t)]^2) \quad \rightarrow$$

$$\frac{\partial S}{\partial t} = \dot{S}_o(t) + \frac{m \ddot{q}(t)}{\hbar} [x - q(t)] - \frac{m \dot{q}(t)^2}{\hbar} +$$

$$+ \frac{m}{2\hbar} [\frac{\ddot{a}(t)}{a(t)} - \frac{\dot{a}^2(t)}{a^2(t)}] [x - q(t)]^2 - \frac{m \dot{q}(t)}{\hbar} \frac{\dot{a}(t)}{a(t)} [x - q(t)]. \quad (3.8)$$

Considering (3.1) let us write $V_{qu}$ given by (2.12a,b) in terms of $[x - q(t)]$. Initially using (2.9b) and (3.1), we calculate the following derivations:

$$\frac{\partial \phi}{\partial x} = \frac{\partial}{\partial x}\left([2\pi a^2(t)]^{-1/4} e^{-\frac{[x-q(t)]^2}{4 a^2(t)}}\right) = [2\pi a^2(t)]^{-1/4} e^{-\frac{[x-q(t)]^2}{4 a^2(t)}} \frac{\partial}{\partial x}\left(-\frac{[x-q(t)]^2}{4 a^2(t)}\right) \rightarrow$$

$$\frac{\partial \phi}{\partial x} = -[2\pi a^2(t)]^{-1/4} e^{-\frac{[x-q(t)]^2}{4 a^2(t)}} \frac{[x-q(t)]}{2 a^2(t)},$$

$$\frac{\partial^2 \phi}{\partial x^2} = \frac{\partial}{\partial x}\left(-[2\pi a^2(t)]^{-1/4} e^{-\frac{[x-q(t)]^2}{4 a^2(t)}} \frac{[x-q(t)]}{2 a^2(t)}\right) =$$

$$= -[2\pi a^2(t)]^{-1/4} e^{-\frac{[x-q(t)]^2}{4 a^2(t)}} \frac{\partial}{\partial x}\left(\frac{[x-q(t)]}{2 a^2(t)}\right) -$$

$$- [2\pi a^2(t)]^{-1/4} e^{-\frac{[x-q(t)]^2}{4 a^2(t)}} \frac{\partial}{\partial x}\left(-\frac{[x-q(t)]^2}{4 a^2(t)}\right) \rightarrow$$

$$\frac{\partial^2 \phi}{\partial x^2} = -[2\pi a^2(t)]^{-1/4} e^{-\frac{[x-q(t)]^2}{4 a^2(t)}} \frac{1}{2 a^2(t)} + [2\pi a^2(t)]^{-1/4} e^{-\frac{[x-q(t)]^2}{4 a^2(t)}} \frac{[x-q(t)]^2}{4 a^4(t)} =$$

$$= -\phi \frac{1}{2 a^2(t)} + \phi \frac{[x-q(t)]^2}{4 a^4(t)} \rightarrow \frac{1}{\phi} \frac{\partial^2 \phi}{\partial x^2} = \frac{[x-q(t)]^2}{4 a^4(t)} - \frac{1}{2 a^2(t)}. \quad (3.9)$$

Substituting (3.9) into (2.12a) and taking into account (3.4), results:

$$V_{qu}(x, t) = V_{qu}[q(t), t] + V_{qu}'[q(t), t] [x - q(t)] + \frac{V_{qu}''[q(t), t]}{2} [x - q(t)]^2 \rightarrow$$

$$V_{qu}(x, t) = \frac{\hbar^2}{4 m a^2(t)} - \frac{\hbar^2}{8 m a^4(t)} [x - q(t)]^2. \quad (3.10)$$

Inserting (2.10a), (3.2), (3.6a,b), (3.8) and (3.10), into (2.14a), we obtain:

$$\hbar \frac{\partial S}{\partial t} + \frac{1}{2} m v_{qu}^2 + 2 \hbar \alpha [S(x, t) - S(x, t - 2L/c)] + V_{qu} =$$

$$= \hbar \left(\dot{S}_o + \frac{m \ddot{q}(t)}{\hbar} [x - q(t)] - \frac{m \dot{q}^2(t)}{\hbar} + \frac{m}{2\hbar}\left[\frac{\ddot{a}(t)}{a(t)} - \frac{\dot{a}^2}{a^2(t)}\right][x - q(t)]^2 -\right.$$

$$- \frac{m \dot{q}(t)}{\hbar} \frac{\dot{a}(t)}{a}(t) [x - q(t)] \left.\right) + \frac{m}{2}\left(\frac{\dot{a}(t)}{a(t)} [x - q(t)] + \dot{q}(t)\right)^2 +$$

$$+ 2 \hbar \alpha \left(S_0(t) + \frac{m \dot{q}(t)}{\hbar} [x - q(t)] + \frac{m}{2\hbar} \frac{\dot{a}(t)}{a(t)} [x - q(t)]^2\right) -$$

$$- 2 \hbar \alpha \left(S_0(t - 2L/c) + \frac{m \dot{q}(t - 2L/c)}{\hbar} [x - q(t)] + \frac{m}{2\hbar} \frac{\dot{a}(t - 2L/c)}{a(t)} [x - q(t)]^2\right) +$$

$$+ \frac{\hbar^2}{4\,m\,a^2(t)} - \frac{\hbar^2}{8\,m\,a^4(t)}\,[x - q(t)]^2 = 0\,. \quad (3.11)$$

Since $(x - q)^o = 1$, expanding (3.11) in potencies of $(x - q)$, we obtain:

$$(\hbar\,\dot{S}_o(t) - m\,\dot{q}(t)^2 + \frac{1}{2} m\,\dot{q}(t)^2 + \frac{\hbar^2}{4\,m\,a(t)^2} + 2\,\hbar\,\alpha\,[\,S_0(t) - S_0(t - 2\,L/c)\,]\,)\,[x - q(t)]^o +$$

$$+ (\,m\,\ddot{q}(t) - m\,\dot{q}(t)\,\frac{\dot{a}(t)}{a(t)} + m\,\dot{q}(t)\,\frac{\dot{a}(t)}{a(t)} + 2\,\alpha\,m\,\dot{q}(t) - 2\,\alpha\,m\,\dot{q}(t - 2\,L/c)\,)\,[x - q(t)] +$$

$$+ (\,\frac{m}{2}\,[\frac{\ddot{a}(t)}{a(t)} - \frac{\dot{a}(t)^2}{a(t)^2}] + \frac{m}{2}\,\frac{\dot{a}^2(t)}{a^2(t)} + m\,\alpha\,\frac{\dot{a}(t)}{a(t)} - m\,\alpha\,\frac{\dot{a}(t - 2\,L/c)}{a(t)} - \frac{\hbar^2}{8\,m\,a^4(t)}\,)\,[x - q(t)]^2 = 0$$

$$. \quad (3.12)$$

As (3.2) is an identically null polynomium, all coefficients of the potencies must be equal to zero, that is:

$$\dot{S}_o(t) = \frac{1}{\hbar}\,(\,\frac{1}{2} m\,\dot{q}(t)^2 - \frac{\hbar^2}{4\,m\,a(t)^2} + 2\,\hbar\,\alpha\,[\,S_0(t - 2\,L/c) - S_0(t)\,]\,)\,, \quad (3.13)$$

$$\ddot{q}(t) - 2\,\alpha\,[\dot{q}(t - 2\,L/c) - \dot{q}(t)] = 0\,, \quad (3.14)$$

$$\ddot{a}(t) - 2\,\alpha\,[\dot{a}(t - 2\,L/c) - \dot{a}(t)] = \frac{\hbar^2}{4\,m^2\,a(t)^3}\,. \quad (3.15)$$

Assuming that the following initial conditions are obeyed:

$$q(0) = x_o\,, \quad \dot{q}(0) = v_o\,, \quad a(0) = a_o\,, \quad \dot{a}(0) = b_o\,, \quad (3.16\text{a-d})$$

and that:

$$S_o(0) = \frac{m\,v_o\,x_o}{\hbar}\,, \quad (3.17)$$

the integration of (3.13) gives:

$$S_o(t) = \frac{1}{\hbar} \int_o^t dt'\,(\,\frac{1}{2} m\,\dot{q}^2(t') + 2\,\alpha\,[\,S_0(t') - S_0(t' - 2\,L/c)\,] -$$

$$- \frac{\hbar^2}{4\,m\,a^2(t')}\,) + \frac{m\,v_o\,x_o}{\hbar}. \quad (3.18)$$

Putting (3.18) into eq. (3.6a) results:

$$S(x,\,t) = \frac{1}{\hbar} \int_o^t dt'\,(\,\frac{1}{2} m\,\dot{q}^2(t') + 2\,\hbar\,\alpha\,[S_0(t') - S_0(t' - 2\,L/c)] - \frac{\hbar^2}{4\,m\,a^2(t')}\,) +$$

$$+ \frac{m v_o x_o}{\hbar} + \frac{m \dot{q}(t)}{\hbar} [x - q(t)] + \frac{m}{2 \hbar} [\frac{\dot{a}(t)}{a(t)} [x - q(t)]^2 ]. \quad (3.19)$$

The above result permit us, finally, to obtain the wave packet for the *SNEEE*. Indeed, considering (2.4a), (2.9b), (3.1) and (3.19), we get: [17]

$$\psi(x, t) = [2 \pi a^2(t)]^{-1/4} exp [ \frac{i m}{2 \hbar} ( \frac{\dot{a}(t)}{a(t)} - \frac{1}{4 a^2(t)} ) [x - q(t)]^2 ] \times$$

$$\times exp [ \frac{i m \dot{q}(t)}{\hbar} [x - q(t)] + \frac{i m v_o x_o}{\hbar} ] \times$$

$$\times exp [ \frac{i}{\hbar} \int_o^t dt' ( \frac{1}{2} m \dot{q}^2(t') + 2 \hbar \alpha [S_0(t') - S_0(t' - 2 L/c)] - \frac{\hbar^2}{4 m a^2(t')} ) ]. \quad (3.20)$$

Note that putting $\alpha = 0$ into (3.20) we obtain the quantum wave packet of the free particle. [2]

### 4. The Feynman-de Broglie-Bohm Propagator of the SNEEE

4.1. Introduction

In 1948, [18] Feynman formulated the following principle of minimum action for the Quantum Mechanics:

*The transition amplitude between the states | a > and | b > of a quantum-mechanical system is given by the sum of the elementary contributions, one for each trajectory passing by | a > at the time $t_a$ and by | b > at the time $t_b$. Each one of these contributions have the same modulus, but its phase is the classical action $S_{c\ell}$ for each trajectory.*

This principle is represented by the following expression known as the "Feynman propagator":

$$K(b, a) = \int_a^b e^{\frac{i}{\hbar} S_{c\ell}(b, a)} D x(t), \quad (4.1)$$

with:

$$S_{c\ell}(b, a) = \int_{t_a}^{t_b} L(x, \dot{x}, t) dt, \quad (4.2)$$

where $L(x, \dot{x}, t)$ is the Lagrangean and $D x(t)$ is the Feynman's Measurement. It indicates that we must perform the integration taking into account all the ways connecting the states | a > and | b >.

Note that the integral which defines $K(b, a)$ is called "path integral" or "Feynman integral" and that the Schrodinger wavefunction $\psi(x, t)$ of any physical system is given by (we indicate the initial position and initial time by $x_o$ and $t_o$, respectively): [19]

$$\psi(x, t) = \int_{-\infty}^{+\infty} K(x, x_o, t, t_o) \psi(x_o, t_o) dx_o, \quad (4.3)$$

with the quantum causality condition:

$$\lim_{t, t_o \to 0} K(x, x_o, t, t_o) = \delta(x - x_o). \quad (4.4)$$

## 4.2. Calculation of the Feynman Propagator of the SNEEE

According to Section 3, the wavefunction $\psi(x, t)$ that was named wave packet of the of the *SNEEE*, can be written as [see (3.20)]:

$$\psi(x, t) = [2\pi a^2(t)]^{-1/4} \exp\left[\frac{i m}{2 \hbar}\left(\frac{\dot{a}(t)}{a(t)} - \frac{1}{4 a^2(t)}\right)[x - q(t)]^2\right] \times$$

$$\times \exp\left[\frac{i m \dot{q}(t)}{\hbar}[x - q(t)] + \frac{i m v_o x_o}{\hbar}\right] \times$$

$$\times \exp\left[\frac{i}{\hbar}\int_o^t dt'\left(\frac{1}{2} m \dot{q}^2(t') + 2 \hbar \alpha [S_0(t') - S_0(t' - 2L/c)] - \frac{\hbar^2}{4 m a^2(t')}\right)\right]. \quad (4.5)$$

where [see (3.14,15)]:

$$\ddot{q}(t) - 2\alpha [\dot{q}(t - 2L/c) - \dot{q}(t)] = 0, \quad (4.6)$$

$$\ddot{a}(t) - 2\alpha [\dot{a}(t - 2L/c) - \dot{a}(t)] = \frac{\hbar^2}{4 m^2 a^{(t)3}}. \quad (4.7)$$

where the following initial conditions were obeyed [see (3.16a-d)]:

$$q(0) = x_o, \quad \dot{q}(0) = v_o, \quad a(0) = a_o, \quad \dot{a}(0) = b_o. \quad (4.8\text{a-d})$$

Therefore, considering (4.3), the Feynman-de Broglie–Bohm propagator will be calculated using (4.5), putting with no loss of generality, $t_o = 0$. Thus:

$$\psi(x, t) = \int_{-\infty}^{+\infty} K(x, x_o, t) \psi(x_o, 0) dx_o. \quad (4.9)$$

Let us initially define the normalized quantity:

$$\Phi(v_o, x, t) = (2\pi a_o^2)^{1/4} \psi(v_o, x, t), \quad (4.10)$$

which satisfies the following completeness relation: [20]

$$\int_{-\infty}^{+\infty} dv_o \Phi^*(v_o, x, t) \Phi(v_o, x', t) = \left(\frac{2\pi\hbar}{m}\right) \delta(x - x'). \quad (4.11)$$

Considering (2.4a), (2.9a) and (4.9), we get:

$$\Phi^*(v_o, x, t) \psi(v_o, x, t) =$$

$$= (2\pi a_o^2)^{1/4} \psi^*(v_o, x, t)\psi(v_o, x, t) = (2\pi a_o^2)^{1/4} \rho(v_o, x, t) \rightarrow$$

$$\rho(v_o, x, t) = (2\pi a_o^2)^{-1/4} \Phi^*(v_o, x, t)\psi(v_o, x, t). \quad (4.12)$$

On the other side, substituting (4.12) into (2.11), integrating the result and using (3.1) and (4.10) results [remembering that $\psi^*\psi(\pm\infty) \rightarrow 0$]: [19]

$$\frac{\partial(\Phi^*\psi)}{\partial t} + \frac{\partial(\Phi^*\psi v_{qu})}{\partial x} = 0 \rightarrow$$

$$\frac{\partial}{\partial t}\int_{-\infty}^{+\infty} dx\, \Phi^*\psi + (\Phi^*\psi v_{qu})|_{-\infty}^{+\infty} =$$

$$= \frac{\partial}{\partial t}\int_{-\infty}^{+\infty} dx\, \Phi^*\psi + (2\pi a_o^2)^{1/4}(\psi^*\psi v_{qu})|_{-\infty}^{+\infty} = 0 \rightarrow$$

$$\frac{\partial}{\partial t}\int_{-\infty}^{+\infty} dx\, \Phi^*\psi = 0. \quad (4.13)$$

Eq. (4.13) shows that the integration is time independent. Consequently:

$$\int_{-\infty}^{+\infty} dx'\, \Phi^*(v_o, x', t)\psi(x', t) = \int_{-\infty}^{+\infty} dx_o\, \Phi^*(v_o, x_o, 0)\psi(x_o, 0). \quad (4.14)$$

Multiplying (4.14) by $\Phi(v_o, x, t)$ and integrating over $v_o$ and using (4.11), we obtain [remembering that $\int_{-\infty}^{+\infty} dx'\, f(x')\delta(x'-x) = f(x)$]:

$$\int_{-\infty}^{+\infty}\int_{-\infty}^{+\infty} dv_o\, dx'\, \Phi(v_o, x, t)\Phi^*(v_o, x', t)\psi(x', t) =$$

$$= \int_{-\infty}^{+\infty}\int_{-\infty}^{+\infty} dv_o\, dx_o\, \Phi(v_o, x, t)\Phi^*(v_o, x_o, 0)\psi(x_o, 0) \rightarrow$$

$$\int_{-\infty}^{+\infty} dx'\, (\frac{2\pi\hbar}{m})\delta(x'-x)\psi(x', t) = (\frac{2\pi\hbar}{m})\psi(x, t) =$$

$$= \int_{-\infty}^{+\infty}\int_{-\infty}^{+\infty} dv_o\, dx_o\, \Phi(v_o, x, t)\Phi^*(v_o, x_o, 0)\psi(x_o, 0) \rightarrow$$

$$\psi(x, t) = \int_{-\infty}^{+\infty}[(\frac{m}{2\pi\hbar})\int_{-\infty}^{+\infty} dv_o\, \Phi(v_o, x, t)\times$$

$$\times \Phi^*(v_o, x_o, 0)] \psi(x_o, 0) \, dx_o. \quad (4.15)$$

Comparing (4.9) and (4.15), we have:

$$K(x, x_o, t) = \frac{m}{2\pi\hbar} \int_{-\infty}^{+\infty} dv_o \, \Phi(v_o, x, t) \Phi^*(v_o, x_o, 0). \quad (4.16)$$

Substituting (4.5) and (4.10) into (4.16), we finally obtain the Feynman – de Broglie–Bohm Propagator of the SNEEE, remembering that

$$\Phi^*(v_o, x_o, 0) = exp\left(-\frac{i m v_o x_o}{\hbar}\right)]:$$

$$K(x, x_o; t) = \frac{m}{2\pi\hbar} \int_{-\infty}^{+\infty} dv_o \sqrt{\frac{a_o}{a(t)}} \times$$

$$\times exp\left[\left(\frac{i m \dot{a}(t)}{2\hbar a(t)} - \frac{1}{4 a^2(t)}\right)[x - q(t)]^2 + \frac{i m \dot{q}(t)}{\hbar}[x - q(t)]\right] \times$$

$$\times exp\left[\frac{i}{\hbar} \int_o^t dt' \left(\frac{1}{2} m \dot{q}^2(t') + 2\hbar\alpha[S_0(t') - S_0(t' - 2L/c)] - \frac{\hbar^2}{4 m a^2(t')}\right)\right], \quad (4.17)$$

where $q(t)$ and $a(t)$ are solutions of the (4.6, 7) differential equations.

Finally, it is important to note that putting $\alpha = 0$ into (3.14), (3.15) and (4.17) we obtain the free particle Feynman propagator. [2, 19]

**NOTES AND REFERENCES**